\documentclass[aps,prb,twocolumn,amsmath,superscriptaddress,amssymb,longbibliography]{revtex4}
\usepackage{graphicx}
\usepackage{enumitem}
\usepackage[colorlinks,bookmarks=false,citecolor=red,linkcolor=blue,urlcolor=blue]{hyperref}

\usepackage{dcolumn}
\usepackage{bm}

\usepackage{amsmath,amssymb}

\usepackage{bbold}

\newcommand{\bea}{\begin{eqnarray}}          
\newcommand{\eea}{\end{eqnarray}}

\allowdisplaybreaks

\begin{document}

\title{Entropic sampling in frustrated magnets: role of self-intersecting spaces}
\author{Alwyn Jose Raja}
\email{alwynjoseraja2000@gmail.com}
\affiliation{Birla Institute of Technology and Science, Pilani 333031, India}
\author{R. Ganesh}
\email{r.ganesh@brocku.ca}
\affiliation{Department of Physics, Brock University, St. Catharines, Ontario L2S 3A1, Canada}

\date{\today}

\begin{abstract}
Frustrated magnets typically possess a large space of classical ground states. If this degeneracy is not protected by symmetry, thermal fluctuations may `select' certain states via order-by-disorder. In this article, we examine a precursor effect where all ground states are sampled, but with different weights. Geometry plays a key role in determining the weight distribution and its behaviour.
We demonstrate this with two examples -- both clusters with four spins coupled by XY interactions. 
In the first, the classical ground states form a smooth space. In the second, they form a self-intersecting non-manifold space. Ground state sampling is very different in these two cases. We first consider the microcanonical ensemble picture, where fluctuations conserve energy. Phase space arguments suggest that the first model exhibits 
energy-independent probabilities. The second shows a dramatic energy-dependence with relative probability increasing as $\epsilon^{-1/2}$, where $\epsilon$ is the energy of the system. We simulate low-energy dynamics in both models, confirming the expected behaviour. 
We next consider the canonical ensemble, where the first model produces temperature-independent probabilities. 
In the second, relative probability rises sharply as $T^{-1/2}$, where $T$ is the temperature. 
Our results bring out a classical analogue of order-by-singularity, a mechanism that has been recently proposed in the context of quantum spin clusters. The sampling of classical orders is qualitatively different in systems with self-intersecting ground state spaces. 
It grows at low energies and becomes singular as $\epsilon \rightarrow 0$ (microcanonical ensemble) or $T\rightarrow 0$ (canonical ensemble). We discuss relevance for disordered phases in macroscopic magnets, particularly for spiral liquids. 
\end{abstract}

\maketitle

\section{Introduction}
A hallmark of frustrated magnetism is large degeneracy of classical ground states. Despite being degenerate, ground states may provide varying scope for fluctuations. 
Originating in quantum or thermal effects, fluctuations lead to differences in zero-point energy and/or free energy. The system settles into the classical ground state with the lowest (free) energy. This state is said to have been `selected' by fluctuations. This phenomenon is well known as `order by disorder'\cite{Villain_1980,Shender_1996}. Recent studies have explored the underlying mechanism by drawing an analogy to particle localization\cite{McClarty2018,Khatua_2021,Khatua_2023}. At low energies, a frustrated magnet can be viewed as a single particle moving on an abstract space -- consisting of all classical ground states. Fluctuations, be they of quantum or thermal origin, give rise to a potential on this space. If the potential is deep enough, the particle localizes at its minimum. This manifests as the magnet settling into one particular classically ordered state.

Classical ordering due to order-by-disorder is analogous to spontaneous symmetry breaking\cite{Goldenfeld,Beekman2019}. The principles of statistical mechanics (the ensemble and ergodic hypotheses) break down so that the system is confined to one out of many degenerate classical ground states. This occurs only at low energy (or low temperature) and large system sizes. In this article, we consider a precursor state where the hypotheses of statistical mechanics hold true. We present arguments in the context of magnetic clusters, where small system size prevents ergodicity-breaking. The same arguments could apply to macroscopic magnets, e.g., above a critical `spontaneous'-ordering temperature. Indeed, recent studies have explored `spiral liquid' phases\cite{Attig_2017,Niggemann_2020,Yao_2021,Yan_2022}  wherein a frustrated magnet simultaneously samples a large set of classical orderings. Our results suggest that such phases may exhibit a fine structure encoded in the relative sampling weights.

A key concept in what follows is the space of classical ground states (CGSS) and its geometric character. 
Various CGSS geometries are known to be realized. Examples include lines\cite{Alexander_1980,Heinila_1993}, circle-like closed curves\cite{Mulder_2010,Balla_2019,Yan_2022}, sheets\cite{Heinila_1993}, surfaces\cite{Attig_2017}, tori\cite{Khatua_2019}, intersecting circles\cite{Srinivasan_2020,Niggemann_2020} and even dense three-dimensional spaces\cite{Iqbal_2019}. In macroscopic magnets, the CGSS is usually described in momentum space using the Luttinger-Tisza approach. In magnetic clusters, the CGSS is often an abstract space arising from geometric constraints.\cite{Chalker2011,Khatua_2019,Srinivasan_2020}. In the context of quantum fluctuations in magnetic clusters, studies (by one of the present authors) have drawn a distinction between two mechanisms that contribute to state selection\cite{Khatua_2021}. 
One is driven by a fluctuation-generated potential that induces localization at its minimum. This is the only possible mechanism in systems where the CGSS is a smooth manifold. 
The second mechanism comes into play if the CGSS self-intersects, e.g., forming a figure-of-eight. It arises from bound state formation at a singularity -- a quantum effect driven by the local topology around an intersection point\cite{He_2023}. 
This mechanism has been termed `order by singularity'\cite{Khatua_2019}. It may have observable consequences, e.g., in the scaling behaviour of the selection-induced energy gap\cite{Khatua_2021}.

Here, we consider CGSS sampling in a purely classical setting. We build upon early work by Moessner and Chalker\cite{MoessnerChalker_1998,Chalker2011}, pointing out contrasting selection behaviour in two magnetic clusters with all-to-all couplings. 
We expand on their findings by contrasting two similar clusters -- one with a smooth CGSS and the other with self-intersections. 
We demonstrate that this difference gives rise to qualitatively different selection behaviour. Unlike the smooth case, self-intersection-driven selection grows dramatically at low energies.

\section{Models}
We discuss two model magnetic clusters below. The first is the symmetric quadrumer, a cluster of four spins with all-to-all XY couplings. It is described by the Hamiltonian
\begin{eqnarray}
H_{sym.} = J \sum_{i<j} \left(\vec{S}_i \cdot \vec{S}_j\right)_\parallel,
\label{Eq.Hsym}
\end{eqnarray}
where the indices $i$ and $j$ run over all pairs chosen from among four spins. The spins are Heisenberg-like with x, y and z components.
However, the couplings are of XY nature, i.e., $(\vec{S}_i \cdot \vec{S}_j)_\parallel \equiv {S}_i^x {S}_j^x + {S}_i^y {S}_j^y$. 
 The coupling constant $J$ is assumed to be positive and is henceforth set to unity. The resulting classical ground states have been discussed in Ref.~\onlinecite{Khatua_2019}. To minimize energy, the four spin vectors must lie in the XY plane \textit{and} add to zero. This can be viewed as orienting the spins as two anti-aligned pairs. This can be done in three distinct ways, each with two free angle variables. This results in a CGSS consisting of three tori. However, the tori are not entirely distinct -- they intersect pairwise along lines that represent collinear states. We discuss a simplified view of this space below.

The second model is the asymmetric quadrumer, also described in Ref.~\onlinecite{Khatua_2019}. It is very similar to the symmetric quadrumer, but with two bonds having a stronger coupling strength. The Hamiltonian is given by 
\begin{eqnarray}
H_{asym.} = H_{sym.} + \lambda \Big\{(\vec S_1 \cdot \vec S_2)_\parallel +(\vec S_3 \cdot \vec S_4)_\parallel \Big\}.
\label{Eq.Hasym}
\end{eqnarray}
Here, $\lambda $ is the anisotropy parameter. A positive value of $\lambda$ forces ground states to have $\vec{S}_1 = - \vec{S}_2$ and $\vec{S}_3 = -\vec{S}_4$, with all four spins lying in the XY plane. The resulting CGSS is a single torus, a space parameterized by two angle variables. 

\begin{figure}
\includegraphics[width=3in]{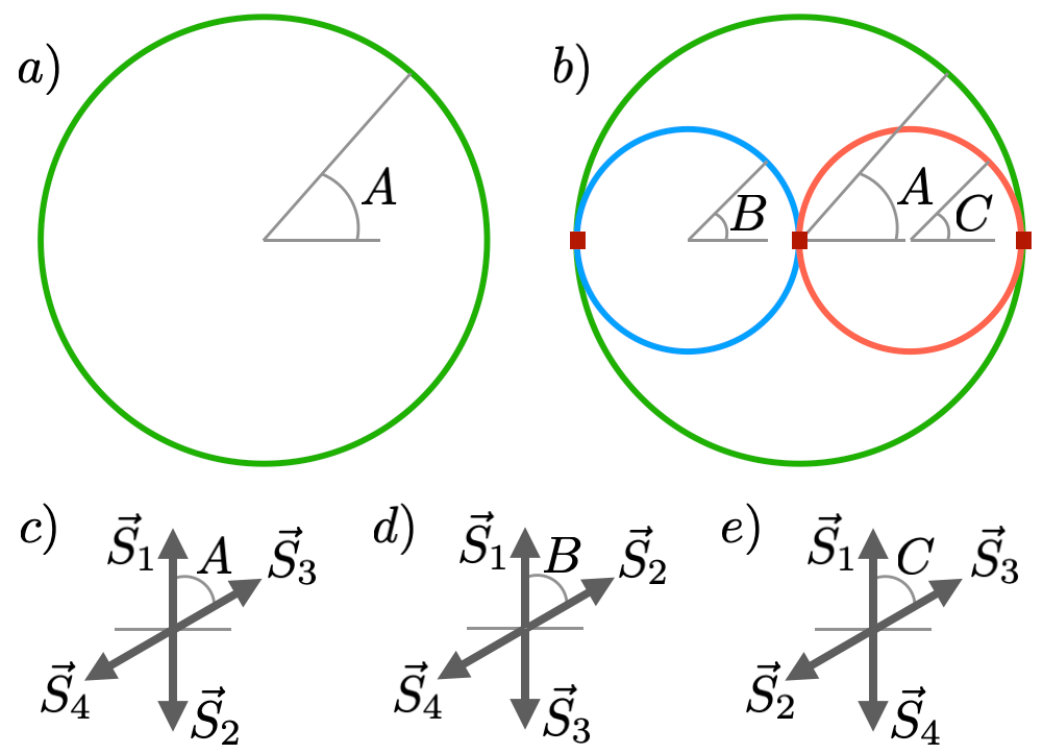}
\caption{Classical ground state spaces (CGSS') of the two models. a) CGSS of the asymmetric quadrumer is a circle, parameterized by one angle $A$. Each point on the circle corresponds to a state as shown in (c) with $\vec{S}_1 = -\vec{S}_2$ and $\vec{S}_3 = -\vec{S}_4$. The angle between $\vec{S}_1$ and $\vec{S}_3$ is denoted as $A$. b) CGSS of the symmetric quadrumer consists of three intersecting circles. The angles $A$, $B$ and $C$ parameterize points on each circle -- they correspond to states shown in (c), (d) and (e). The intersection points are marked with red squares. All spins lie within the XY plane. All states allow for a global rotation about the Z axis.    }
\label{fig.cartoon}
\end{figure}

In both models, the CGSS is larger than the space of Hamiltonian symmetries. The only continuous Hamiltonian symmetry is global rotation about the spin-$Z$ axis. With this symmetry in mind, we work in a frame that co-rotates with $\vec{S}_1$. The first spin, $\vec{S}_1$ can now be taken to have a fixed orientation, say along the Y axis. 
The CGSS of the asymmetric quadrumer in this frame is shown in Fig.~\ref{fig.cartoon}(a). 
It is a circle parameterized by a single angle $A$, the angular displacement between $\vec{S}_1$ and $\vec{S}_3$. The other two spins are immediately determined as $\vec{S}_2 = -\vec{S}_1$ and $\vec{S}_4 = -\vec{S}_3$. 
The CGSS of the symmetric quadrumer in this frame is shown in Fig.~\ref{fig.cartoon}(b). It consists of three circles. The circle parameterized by the angle $A$ is the same as that in the asymmetric quadrumer case. The circle parameterized by $B$ has $\vec{S}_3 = -\vec{S}_1$ and $\vec{S}_4 =-\vec{S}_2$, with $B$ representing the relative angle between $\vec{S}_1$ and $\vec{S	}_2$. Similarly, the third circle has $\vec{S}_4 = -\vec{S}_1$ and $\vec{S}_3 =-\vec{S}_2$, with $C$ representing the relative angle between $\vec{S}_1$ and $\vec{S}_3$.

In the symmetric quadrumer, the three circles of the CGSS intersect pairwise. These points of intersection are collinear states. For example, the circles parameterized by $A$ and $B$ share a common point, where $\vec{S}_1 = \vec{S}_4 = -\vec{S}_2 = -\vec{S}_3$. 
As we show below, these collinear states are favoured by fluctuations.

\section{Phase space in the microcanonical ensemble}
To describe state selection, we consider the low-energy behaviour of the two models. We first adopt the microcanonical approach where the total energy is held fixed. We restrict our attention to low energies, i.e., where the total energy is only slightly higher than the classical ground state energy. We describe the resulting phase space and sampling probabilities.    

\subsection{Asymmetric quadrumer}
\label{ssec.asym}
Any classical ground state of the asymmetric quadrumer can be written as 
\begin{eqnarray}
    \vec{S}_1 = -\vec{S}_2 =\hat{n}(\phi_1);~~   \vec{S}_3 = -\vec{S}_4 =\hat{n}(\phi_2),
    \label{eq.circle1}
\end{eqnarray}
where the length of each spin is taken to be unity. We have defined $\hat{n}(\phi) = \cos \phi ~\hat{x} + \sin\phi ~\hat{y} $. This state can be viewed as two rods within the plane, one corresponding to $\vec S_{1,2}$ and other to $\vec{S}_{3,4}$. In the frame that co-rotates with $\vec{S}_1$ (see Fig.~\ref{fig.cartoon}), $\phi_1$ is fixed. Choosing a value for $\phi_2$ amounts to picking one particular classical ground state. 
Low energy states may have small deviations away from this state. 
 A generic low-energy state can be expressed in terms of six fluctuation variables $\{ \ell_1, \ell_2, \mu_1, \mu_2, m_1, m_2\}$ as follows:
\begin{eqnarray}
\nonumber     \vec S_1 &=& \hat n(\phi _1)+ \ell_1~ \hat{\zeta}(\phi_1)+ \mu_1~ \hat{z} + m_1~\hat z;\\
\nonumber     \vec S_2 &=& -\hat n(\phi _1)+ \ell_1~ \hat{\zeta}(\phi_1)+ \mu_1~ \hat{z} - m_1~\hat z;\\
  \nonumber  \vec S_3 &=& \hat n(\phi _2)+ \ell_2~\hat{\zeta}(\phi_2) + \mu_2 ~\hat{z}+ m_2~\hat z;\\
     \vec S_4 &=& -\hat n(\phi _2)+ \ell_2 ~\hat{\zeta}(\phi_2)+  \mu_2~ \hat{z}- m_2~\hat z.
\label{eq.forms}
\end{eqnarray}
Here, $\mu_1$ and $\mu_2$ represent out-of-plane buckling of the rods. In contrast, $m_1$ and $m_2$ represent out-of-plane tilting fluctuations. The  components $l_1$ and $ l_2$ represent buckling deformations within the plane. We have defined unit vectors denoted by $\hat{\zeta}$ that are perpendicular to $\hat{n}$'s, with $\hat{\zeta}(\phi) = \sin\phi ~\hat{x}  - \cos \phi ~\hat{y}$. The expressions in Eqs.~\ref{eq.forms} include all possible length-preserving deformations. It can be easily checked that $\vert \vec{S}_j \vert^2$ is unity ($j=1,2,3,4$), up to corrections that are quadratic in the fluctuation variables.

Having parameterized low-energy states, we consider the low-energy phase space. 
The Hamiltonian of Eq.~\ref{Eq.Hasym} reduces to
\begin{eqnarray}
\nonumber H_{asym.} &\approx& E_{CGS}+ (1+\lambda)(\mu_1^2+m_1^2+\mu_2^2+m_2^2) \\ 
&+& (1+\lambda +\cos A)~\ell_+^2+(1+\lambda -\cos A)~\ell_-^2.~~~
 \label{eq.Hasymfluc}
\end{eqnarray} 
The `$\approx$' symbol indicates that this form is valid for small fluctuations, where terms beyond quadratic order can be neglected. Here, $E_{CGS} = -2(1+\lambda) $ is the classical ground state energy. The variable $A = \phi_2 - \phi_1$ (see Fig.~\ref{fig.cartoon}) identifies a point on the CGSS. It encodes the classical ground state that is closest. We have defined $\ell_{\pm} = 2(\ell_1 \pm \ell_2)$.

In Eq.~\ref{eq.Hasymfluc}, we have arrived at an expression for the energy of low-lying states. As it contains six quadratic terms, a constant-energy-surface is an ellipsoid in six dimensions. We now take the microcanonical view by demanding that the energy must lie within $(E_{CGS} + \epsilon, E_{CGS} + \epsilon + d\epsilon)$, where $d\epsilon \ll \epsilon \ll 1$. The low energy condition dictates that the energy, $\epsilon$, must be much smaller than the coupling constant (set to unity) which sets the energy scale in the problem. The microcanonical approach allows for a small energy spread $d\epsilon$ that is much smaller than the energy content, $\epsilon$. We arrive at a region of the six-dimensional phase space -- an ellipsoidal shell. We seek to find the volume of this `accessible' region. As $d\epsilon$ is small, this volume can be expressed as $V_\epsilon^{\epsilon+d\epsilon} d\epsilon$. Going further, we wish to resolve this volume into neighbourhoods around each point in the CGSS. To do so, we express the accessible volume as 
\begin{eqnarray}
V_\epsilon^{\epsilon+d\epsilon} d\epsilon \sim  \left\{\int d A~~ v(\epsilon,A) \right\} d\epsilon,
\label{eq.Vphi12}
\end{eqnarray}
where $A$ is the angle that parameterizes the CGSS. 
Relegating details to the appendix, we find
\begin{eqnarray}
v(\epsilon,A) = 
\frac{\pi ^3}{6} \frac{32}{(\lambda +1)^2\sqrt{(1+\lambda)^2- \cos^2 A}} ~~3 \epsilon^2 .
\label{eq.vephi}
\end{eqnarray}
We now assert that the principles of statistical mechanics, viz. the ergodic and equiprobability hypotheses, hold true. They dictate that the probability of a certain classical ground state being sampled is proportional to the volume of the accessible phase space in its neighbourhood, i.e., 
\begin{eqnarray}
P(A) \sim v(\epsilon,A) \sim \frac{1}{\sqrt{(1+\lambda)^2- \cos^2 A}}.
\label{eq.Pasym}
\end{eqnarray}
From the expression for $v(\epsilon,A)$ in Eq.~\ref{eq.vephi} above, we see that the probability for any $A$ depends on the energy via an overall factor of $\epsilon^2$. 
If we compare two values $A_1$ and $A_2$, we find that $P(A_1)/P(A_2)$ is independent of $\epsilon$. 
This results in an \textit{energy-independent} distribution of probabilities over the CGSS, as reflected in Eq.~\ref{eq.Pasym}.

\subsection{Symmetric quadrumer}
\label{ssec.sym}
Starting from the asymmetric quadrumer, we obtain the symmetric quadrumer by tuning the anisotropy parameter, $\lambda$, to zero. For any positive value of $\lambda$, the CGSS is a circle. Precisely at $\lambda = 0$, the CGSS expands to form three intersecting circles as shown in Fig.~\ref{fig.cartoon}. For simplicity, we restrict our attention to points on one of the circles when characterizing the phase space. Similar arguments hold on the other two circles. 

We consider states characterized by Eq.~\ref{eq.circle1} above, shown in Fig.~\ref{fig.cartoon}(a). Although they were introduced as classical ground states of the asymmetric quadrumer, they are ground states of the symmetric quadrumer as well. They correspond to one of the circles in the CGSS. 

Following the same arguments as in the asymmetric quadrumer, low-energy states can be parameterized as in Eq.~\ref{eq.forms}. Their energy takes the same form as given in Eq.~\ref{eq.Hasymfluc}, with six quadratic terms. However, a crucial difference emerges in the $\lambda \rightarrow 0$ limit. Consider the quadratic term proportional to $\ell_+^2$. Its prefactor, given by $1 + \cos A $, vanishes when $A \rightarrow \pi$. Likewise, the prefactor of the $\ell_-^2$ term vanishes when $A \rightarrow 0$. This reflects the change in CGSS topology as $\lambda \rightarrow 0$. For a generic value of $A$, the low-energy Hamiltonian has six quadratic terms. However, at two special values of $A$, the Hamiltonian reduces to five quadratic terms. These points represent collinear states where the CGSS self-intersects. The vanishing of one quadratic coefficient can be viewed as mode softening. At these collinear states, the system may leave one circle of the CGSS and move to another (see Fig.~\ref{fig.cartoon}). 

This has dramatic consequences for the phase space volume. In the vicinity of a non-collinear state, arguments from Sec.~\ref{ssec.asym} continue to hold. The phase space $v(\epsilon, A)$ takes the same form as in Eq.~\ref{eq.vephi} above. The probability of the state being sampled is $P(\epsilon,\mathrm{non}-\mathrm{collinear})\sim \epsilon^2$. However, when $A = 0$ or $\pi$, the phase space is qualitatively different. It constitutes an ellipsoid in five dimensions, with a free sixth coordinate. This coordinate can be freely integrated over its range, as it is not constrained by energy -- this will contribute an energy-independent factor to the phase space volume. Based on these arguments, we may write 
\begin{eqnarray}
v(\epsilon,\mathrm{collinear}) 
\sim \epsilon^{3/2}.
\end{eqnarray}
We have used the fact the volume of an ellipsoidal shell in $D$ dimensions scales as $R^{D-1}dR$, where $R$ is the length scale (say, the semi-major axis). Here, we have $D=5$ and $R \sim \sqrt{\epsilon}$ so that the volume scales as $ \epsilon^{3/2} d\epsilon$. Crucially, the volume scales as a different power of energy when compared with non-collinear states. This leads to a stark difference when comparing probabilities,  
\begin{eqnarray}
\frac{P(\epsilon,\mathrm{collinear})}{P(\epsilon,\mathrm{non-collinear})} \sim \frac{\epsilon^{3/2}}{\epsilon^2} \sim \epsilon^{-1/2}.
\end{eqnarray}
Unlike the asymmetric quadrumer, relative probability depends on energy. In fact, it varies dramatically when $\epsilon \rightarrow 0$, as collinear states dominate over all others.

\section{Probabilities from energy-preserving dynamics}
In the previous section, we have examined relative probabilities over the CGSS from phase space considerations. The asymmetric quadrumer yields energy-independent relative probabilities. In contrast, in the symmetric quadrumer, the weight of collinear states grows sharply at low energies. We now verify these results by numerical simulations of low-energy spin-dynamics. As initial conditions for our simulations, we generate random configurations that lie within a suitable energy window.
As the system evolves in time, we track the amount of time spent in the vicinity of each classical ground state. We interpret this as a probability distribution over the CGSS. 

The time-evolution of classical magnets is described by the Landau-Lifshitz equation\cite{Landau_1935}, written succinctly as  
\begin{eqnarray}  
\frac{d\vec{S}_j}{ dt } = \vec{S}_j \times \vec{B}_{\mathrm{eff},j}.
\label{eq.LL}
\end{eqnarray}
The vector $\vec{B}_{\mathrm{eff},j}$ represents the effective magnetic field seen by the $j^\mathrm{th}$ spin. Its form is obtained by re-expressing the Hamiltonian as $H =-\vec{B}_{\mathrm{eff},j} \cdot \vec{S}_j$, where $j$ runs over the four spins of the cluster. For any given spin, the effective field is given in terms of all other spins.  
Eq.~\ref{eq.LL} encodes a set of twelve coupled ordinary differential equations as we have four spins ($j=1,2,3,4$) with three components each. Given an initial condition, the time evolution of the system can be found by numerical integration by Runge-Kutta methods. Up to numerical errors, Eq.~\ref{eq.LL} preserves spin lengths as well as the energy. 

We consider the symmetric and asymmetric quadrumers within the microcanonical ensemble. We fix a small energy window $(E_{CGS}+\epsilon,E_{CGS}+\epsilon+d\epsilon)$ where $E_{CGS} = -2(1+\lambda) $ is the classical ground state energy.

\subsection{Dynamics of the asymmetric quadrumer}
For each value of $\epsilon$, we choose an energy interval $d\epsilon  = 2\times10^{-3} \epsilon$. 
We generate 3 $\times$ 10$^4$ initial configurations with energy within this window. Each configuration is time evolved for $10^4$ time units (time is measured in units of $J^{-1}$, where $J$ is the coupling constant -- set to unity). As time evolution proceeds, the low value of energy guarantees that the system will always be in the vicinity of the CGSS. At each time, we determine the closest point on the CGSS, i.e., the classical ground state that is closest to the current configuration, see Appendix~\ref{app.nearest}. To characterize sampling probability, we divide the CGSS into bins of width one degree (note that the CGSS is parameterized by angle variables). We interpret the fraction of time spent in each interval as the probability of sampling that neighbourhood of the CGSS.

\begin{figure*}
\includegraphics[width=2.3in]{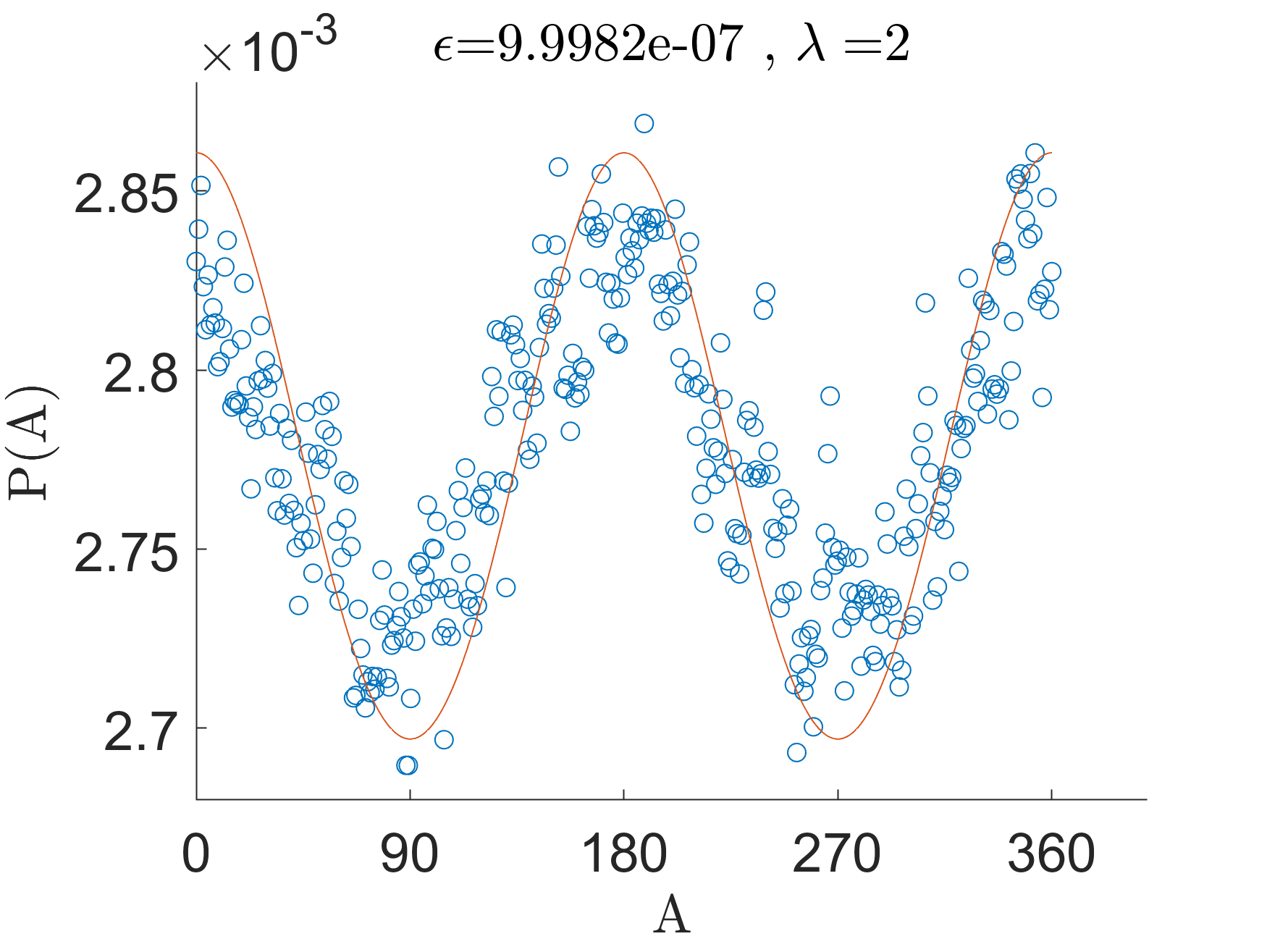}
\includegraphics[width=2.3in]{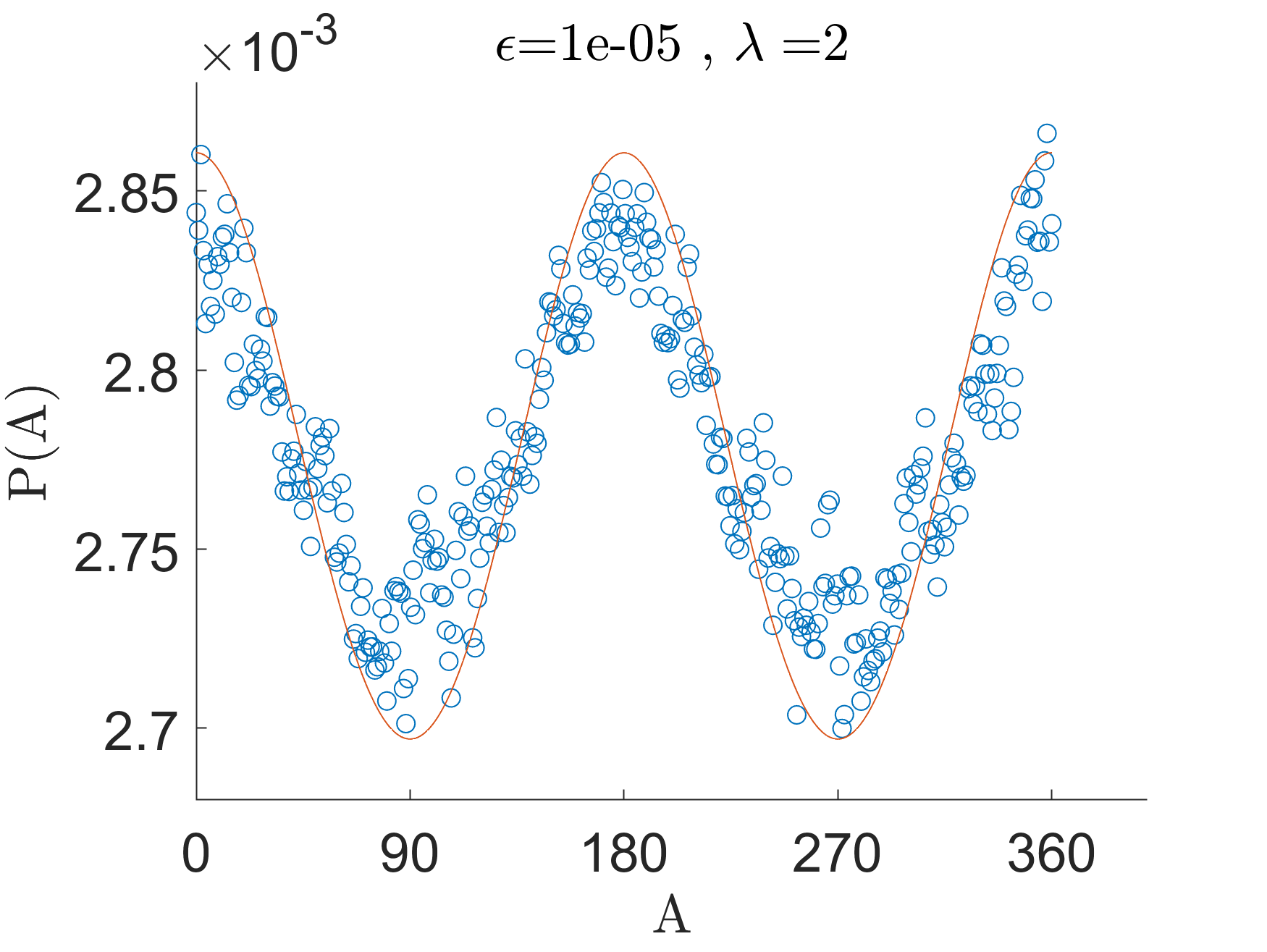}
\includegraphics[width=2.3in]{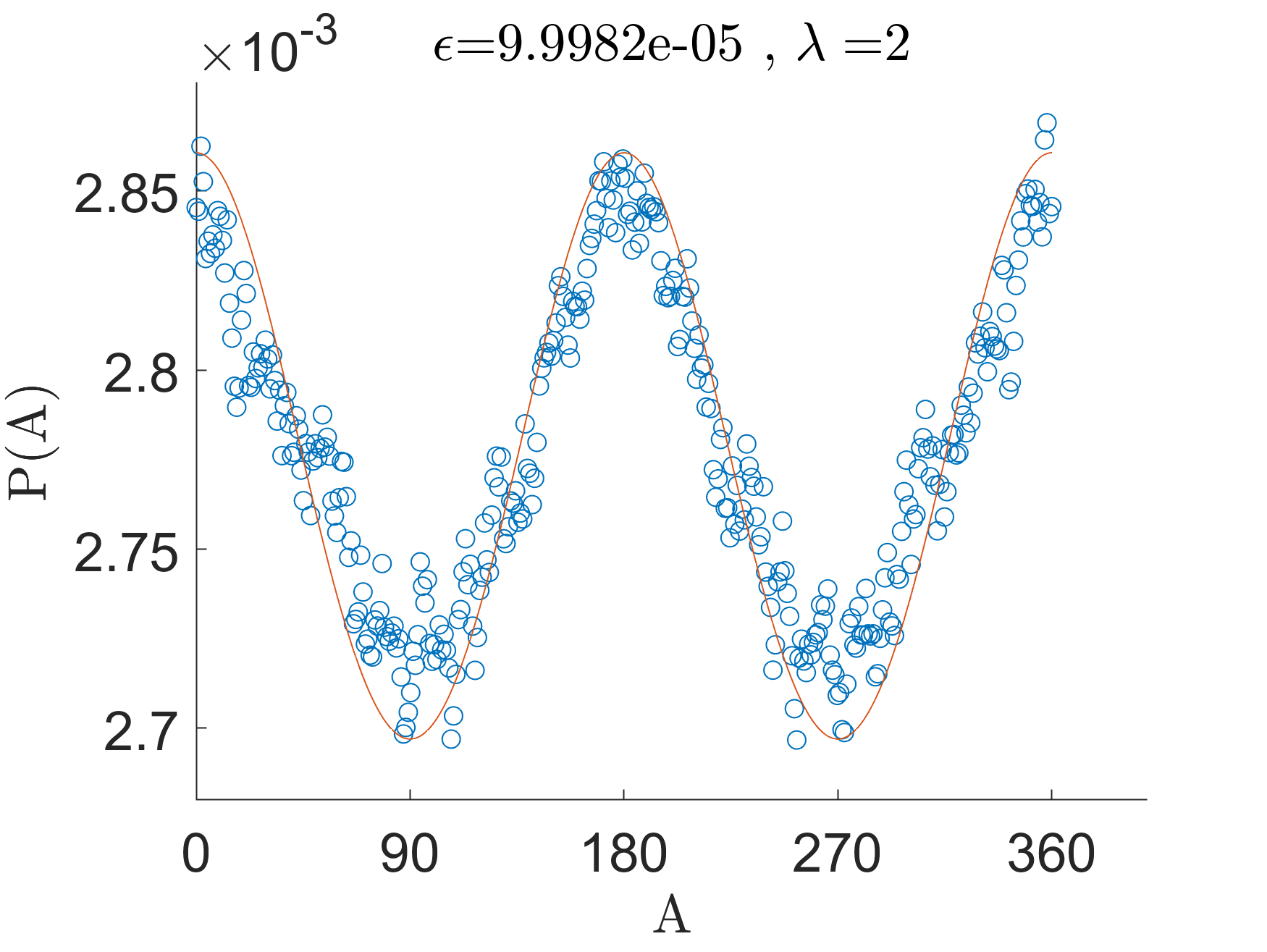}
\caption{Sampling probability over the CGSS coordinate $A$ (in degrees) in the asymmetric quadrumer with $\lambda=2$. Plots correspond to energies $\epsilon\sim 10^{-6},10^{-5},10^{-4}$ (left to right). Data are obtained from energy-conserving Landau-Lifshitz dynamics. In all three panels, they follow the same curve -- obtained from Eq.~\ref{eq.Pasym} with $\lambda=2$ and no fitting parameters.  } 
\label{fig.asymLL}
\end{figure*}

Fig.~\ref{fig.asymLL} shows the obtained probability distributions. The asymmetry parameter is set at $\lambda=2$. Note that collinear states ($A=0$ or $\pi$) have the highest probability -- they are selected by fluctuations. The figure shows data for three values of $\epsilon$. In all three, the data follow the same curve -- that given by Eq.~\ref{eq.Pasym}. The probabilities follow the same form even as the energy is varied over three orders of magnitude. This is in line with the arguments of Sec.~\ref{ssec.asym} where ground-state sampling  
was argued to be energy-independent.

Fig.~\ref{fig.asymLL} depicts state selection for $\lambda = 2$, a strong value of the anisotropy parameter. For weaker values of $\lambda$, the numerical simulations deviate from the expected probability distribution. We believe this is tied to the presence of a large number of periodic orbits. With most initial conditions, the system evolves in a periodic fashion. As a result, it may not sample the accessible phase space in a uniform manner. 

 \subsection{Dynamics of the symmetric quadrumer}
\begin{figure*}
\includegraphics[width=2.3in]{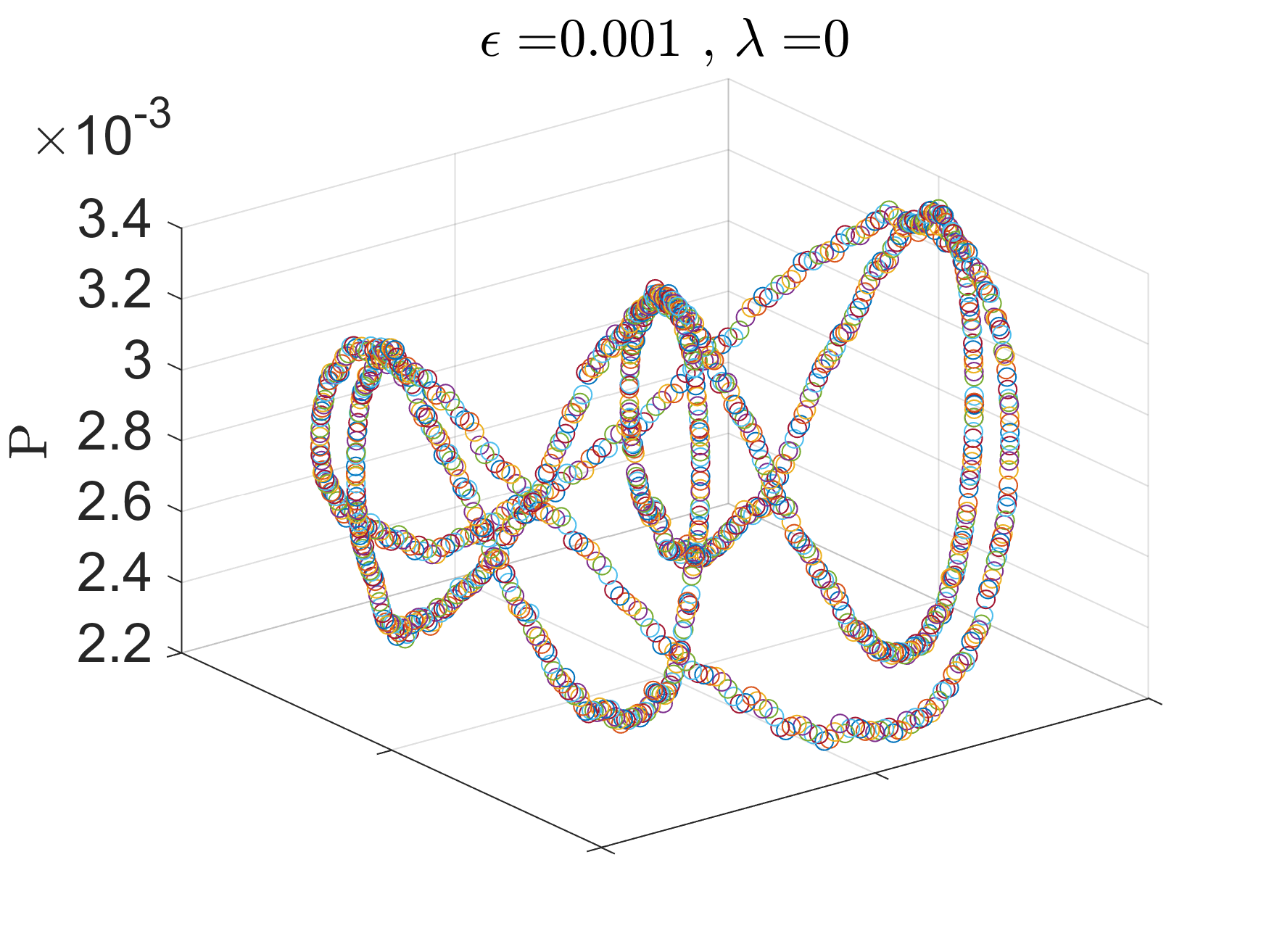}
\includegraphics[width=2.3in]{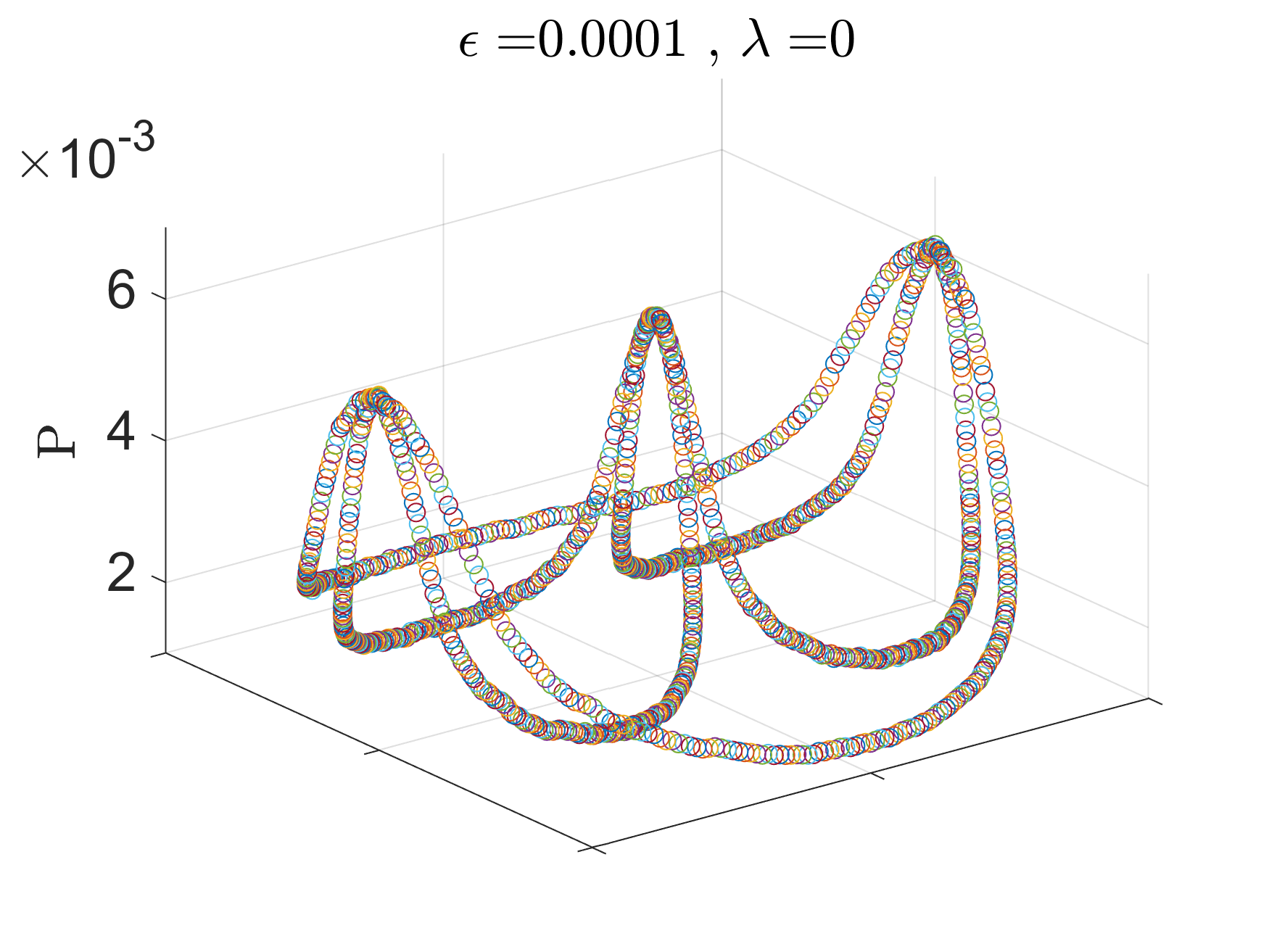}
\includegraphics[width=2.3in]{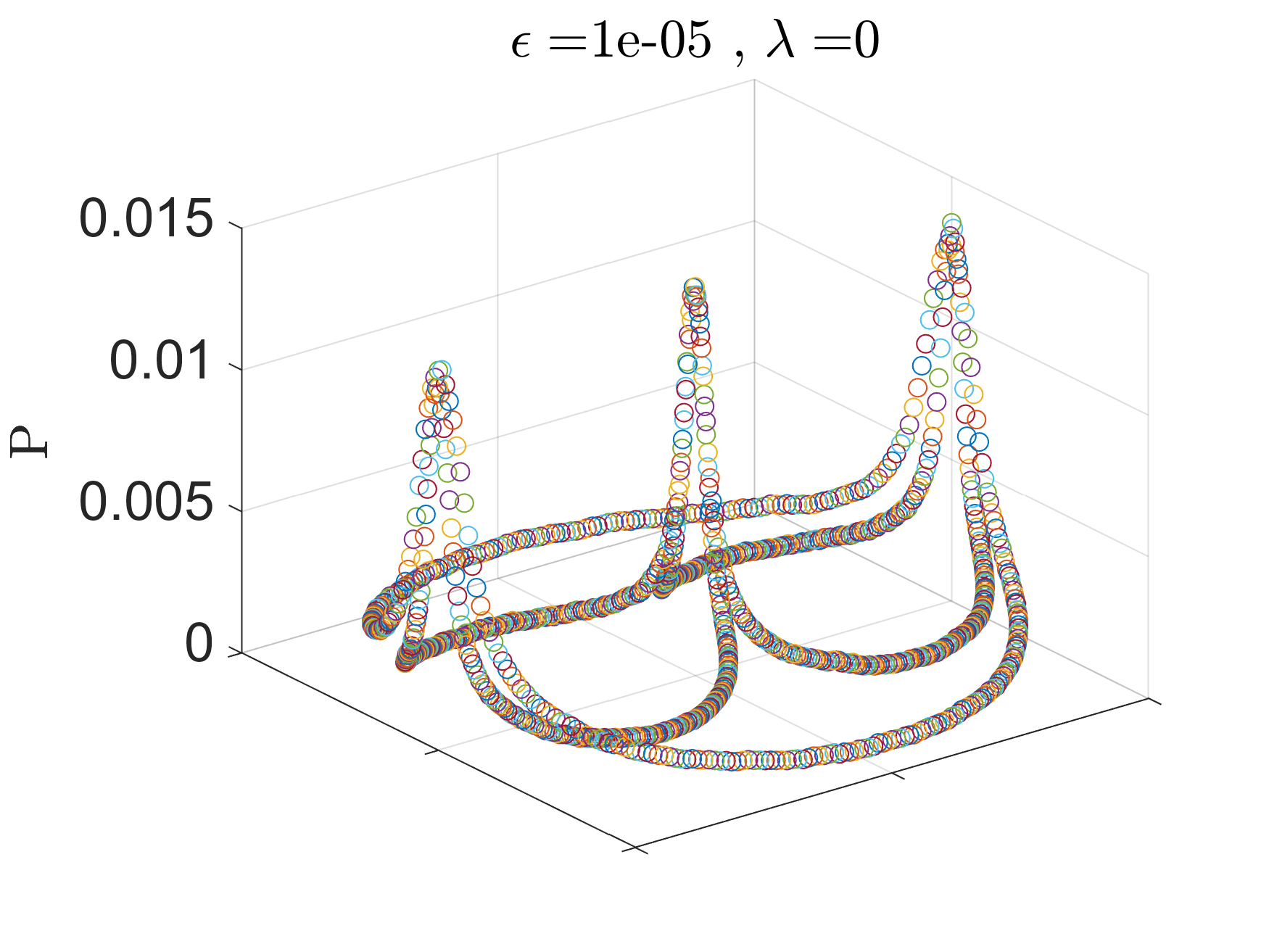}
\caption{Sampling probability in the symmetric quadrumer, obtained from energy-conserving Landau-Lifshitz dynamics. Plots correspond to energies $\epsilon=10^{-3},10^{-4}$ and $10^{-5}$ (left to right). Seen from the top, there are three circles as shown in the CGSS depicted in Fig.~\ref{fig.cartoon}(b). The z-axis represents probability; note that the panels have different z-axis ranges.} 
\label{fig.symLL}
\end{figure*}

 We simulate time evolution in the same manner for the symmetric quadrumer. We vary energy, $\epsilon$, over a large range and fix $d \epsilon = 2\times 10^{-3}\epsilon$.  For each energy, we generate 2000 samples as initial conditions for time evolution. We simulate time evolution for 5 $\times$ 10$^{4}$ time units.  
As the CGSS consists of three intersecting circles, we track the nearest point on this space. Fig.~\ref{fig.symLL} shows the obtained probability distribution over these three circles. This is shown for three different energies. In all three, the probability is highest at the intersection points on the CGSS (collinear states). Unlike the asymmetric quadrumer, the resulting curves vary dramatically with energy. As energy decreases, the probability curves become more sharply peaked. The selection of collinear states strengthens with decreasing energy ($\epsilon$). 
 
\begin{figure}
\includegraphics[width=\columnwidth]{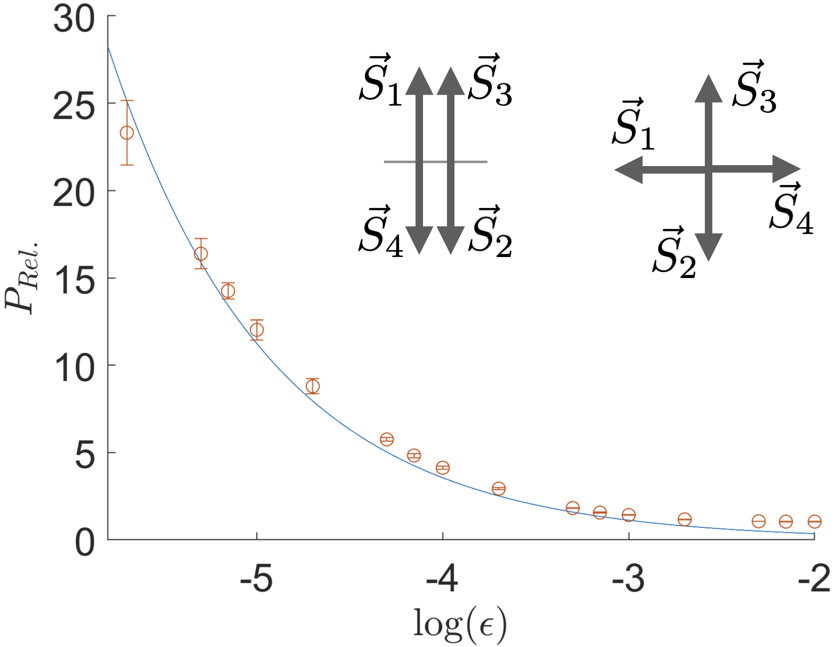}
\caption{Relative probability of collinear and perpendicular states in the symmetric quadrumer vs. energy. The x-axis represents $\log (\epsilon)$ (logarithm with base 10). Error bars are estimated from the spread in values with various choices of collinear and perpendicular states. The data are fit to a curve $P_{rel.} \sim\epsilon^{-1/2}$. The configurations shown are examples of collinear and perpendicular states on the CGSS. } 
\label{fig.symrelprob}
\end{figure}

 Fig.~\ref{fig.symrelprob} compares the relative probability between collinear states and perpendicular states. The latter are states where the four spins point toward the corners of a square -- as seen from Fig.~\ref{fig.symLL}, these perpendicular states have the lowest probability on the CGSS. Fig.~\ref{fig.symrelprob} shows that the relative probability varies strongly with energy. It fits well to $\sim \epsilon^{-1/2}$ as predicted by phase space arguments in Sec.~\ref{ssec.sym}.
 
\section{Probabilities in the canonical ensemble}
 \label{sec.canonical}
 When coupled to a reservoir, the energy of a system will vary with time. Phase space is sampled according to the temperature, a property of the reservoir. We may write the partition function as 
 \begin{eqnarray}
\mathcal{Z}_{canonical} = \int e^{-  \epsilon/T} V_\epsilon^{\epsilon + d\epsilon} d\epsilon,
\label{eq.Zcanonical}
\end{eqnarray}
where $T$ is the temperature, measured in units of energy (so that the Boltzmann constant is unity). The quantity $V_\epsilon^{\epsilon + d\epsilon} d\epsilon$ represents the volume of phase space that lies within an energy window $(\epsilon,\epsilon+d\epsilon)$. This is precisely the volume that was evaluated in the context of the microcanonical ensemble. 

At low energies, the volume $V_\epsilon^{\epsilon+d\epsilon}$ involves an integral over all phase space coordinates. As with Eq.~\ref{eq.Vphi12} above, we separate out the coordinate that parameterizes the CGSS. For the asymmetric quadrumer, we write
\begin{eqnarray}
\mathcal{Z}_{canonical}^{asym.} =\int d A~{z}_{canonical}^{asym.} (A), 
\label{eq.Zcanseparated}
\end{eqnarray}
where $A$ is the CGSS coordinate. We have
\begin{eqnarray}
{z}_{canonical}^{asym.} (A)= \int  e^{-  \epsilon/T} v (\epsilon, A) d\epsilon.
\end{eqnarray}
As seen from Eq.~\ref{eq.vephi} above, $v (\epsilon, A)\sim \epsilon^2$. As a result, the integral over $\epsilon$ can be carried out in a straightforward fashion. We have
\begin{eqnarray}
 {z}_{canonical}^{asym.} (A) \sim f(A) \int_0^\infty d\epsilon ~e^{-\epsilon/T}~\epsilon^2 
  \sim {f(A)}~{T^3} .~~
\end{eqnarray}
Here, $f(A)$ is some function that depends on $A$ but not on $\epsilon$. The limits in the $\epsilon$ integral are taken to be $[0,\infty)$. As $\epsilon$ represents energy cost over the classical ground state ($\epsilon = E_{system} - E_{CGS}$), the lowest value it can take is zero. As contributions from high energy states are exponentially suppressed, we may safely extend the integration to $\epsilon \rightarrow \infty$. At the final step, we have extracted the temperature dependence. While the integral can be evaluated explicitly, the $T$-dependence can be extracted immediately on dimensional grounds.

We now assert that given a point on the CGSS labelled by $A$, the probability of the system exploring its neighbourhood is proportional to ${z}_{canonical}^{asym.} (A)$. From the expression above, we see that the temperature-dependence in ${z}_{canonical}^{asym.} (A)$ comes from an overall factor of $T^3$. Given two points on the CGSS, $A_1$ and $A_2$, the ratio $P(A_1)/P(A_2)$ will not depend on $T$.
As a result, the relative probability between any two points on the CGSS is \textit{temperature-independent}.

We now consider Eq.~\ref{eq.Zcanonical} for the symmetric quadrumer. We resolve this integral into neighbourhoods around each point of the CGSS. On each of the three circles of the CGSS, we obtain an equation of the same form as Eq.~\ref{eq.Zcanseparated} above. However, as argued in Sec.~\ref{ssec.sym} above, the $\epsilon$-dependence of $v (\epsilon, A) $ varies with $A$. At collinear points (i.e., for $A = 0$ or $\pi$), we have $ v (\epsilon, A)  \sim \epsilon^{3/2}$. At all other points, $ v (\epsilon, A)  \sim \epsilon^{2}$. We carry out the $\epsilon$-integral separately for the two cases. We have
\begin{eqnarray}
{z}_{canonical}^{sym.} (A=0,\pi) \sim \int  e^{-  \epsilon/T} \epsilon^{3/2}d\epsilon \sim T^{5/2}.
\end{eqnarray}
In contrast, 
\begin{eqnarray}
{z}_{canonical}^{sym.} (A\neq 0,\pi) \sim \int  e^{-  \epsilon/T} \epsilon^{2}d\epsilon \sim T^{3}.
\end{eqnarray}
We now identify ${z}_{canonical}^{sym.} (A=0,\pi) $ as the probability of a collinear state. The probability of accessing a non-collinear classical ground state is ${z}_{canonical}^{sym.} (A\neq 0,\pi) $. These two probabilities scale differently with temperature, so that 
\begin{eqnarray}
\frac{P(T,\mathrm{collinear})}{P(T,\mathrm{non-collinear})} \sim \frac{T^{5/2}}{T^3} \sim T^{-1/2}.
\end{eqnarray}
That is, the relative probability between collinear and non-collinear states grows as temperature is lowered. In the $T\rightarrow 0$ limit, collinear states dominate. 

We conclude that sampling-by-thermal-fluctuations is qualitatively different between the two models. In the asymmetric quadrumer, the sampling is temperature-independent. In the symmetric quadrumer, sampling-bias grows as temperature decreases. In fact, the domination of collinear states is perfect in the $T\rightarrow 0$ limit.  

\section{Monte Carlo simulations}
We now verify the arguments of the previous section by explicitly simulating thermal fluctuations. For each cluster, we carry out Monte Carlo simulations\cite{Landau_Binder_2014} at various temperatures. 
Single-spin Metropolis moves lead to very low acceptance as they invariably take the system away from the ground state space. Therefore, we employ all-spin moves where each spin is simultaneously deflected from its orientation by an angle $\Delta(T)$. The direction of the deflection is chosen at random. The angle $\Delta(T)$ is varied between 0.2$^\circ$ and 1$^\circ$to ensure an acceptance rate of $\sim 20\%$. For each temperature, we carry out $500-3000$ runs, each consisting of $10^6-10^7$ moves. 

From the Monte Carlo simulations, we extract the sampling probability over the CGSS. We divide the CGSS circle(s) into 360 bins, each of width 1$^\circ$. At each Monte Carlo time, we identify the classical ground state that is closest to the current configuration, see Appendix~\ref{app.nearest}. We assign this to one of the 360 intervals. As the simulation proceeds, we keep track of the amount of Monte Carlo time spent within each bin. The fraction of time spent in each bin is interpreted as the probability of sampling that neighbourhood of the CGSS.

\begin{figure*}
 \includegraphics[width=2.3in]{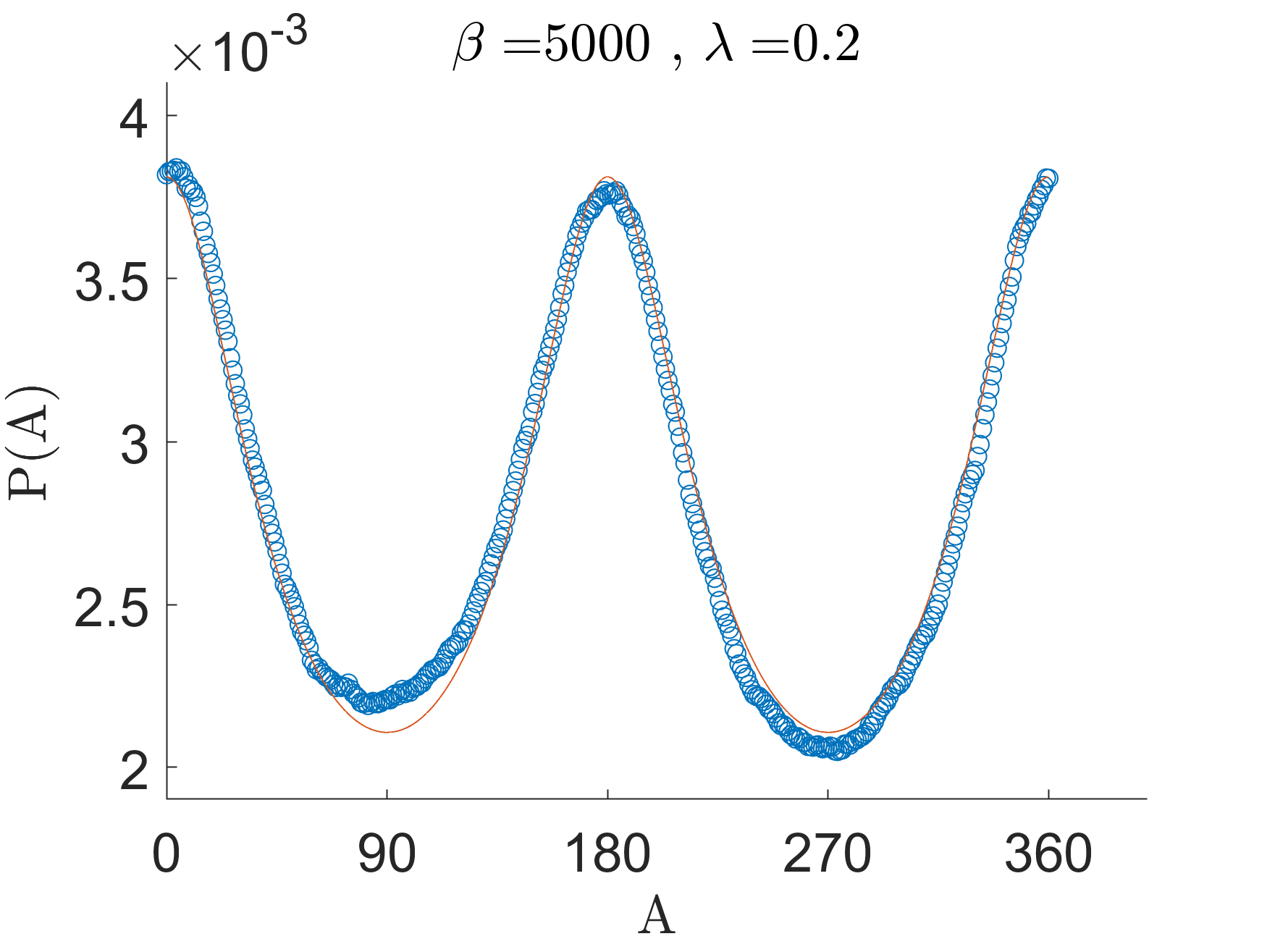}
    \includegraphics[width=2.3in]{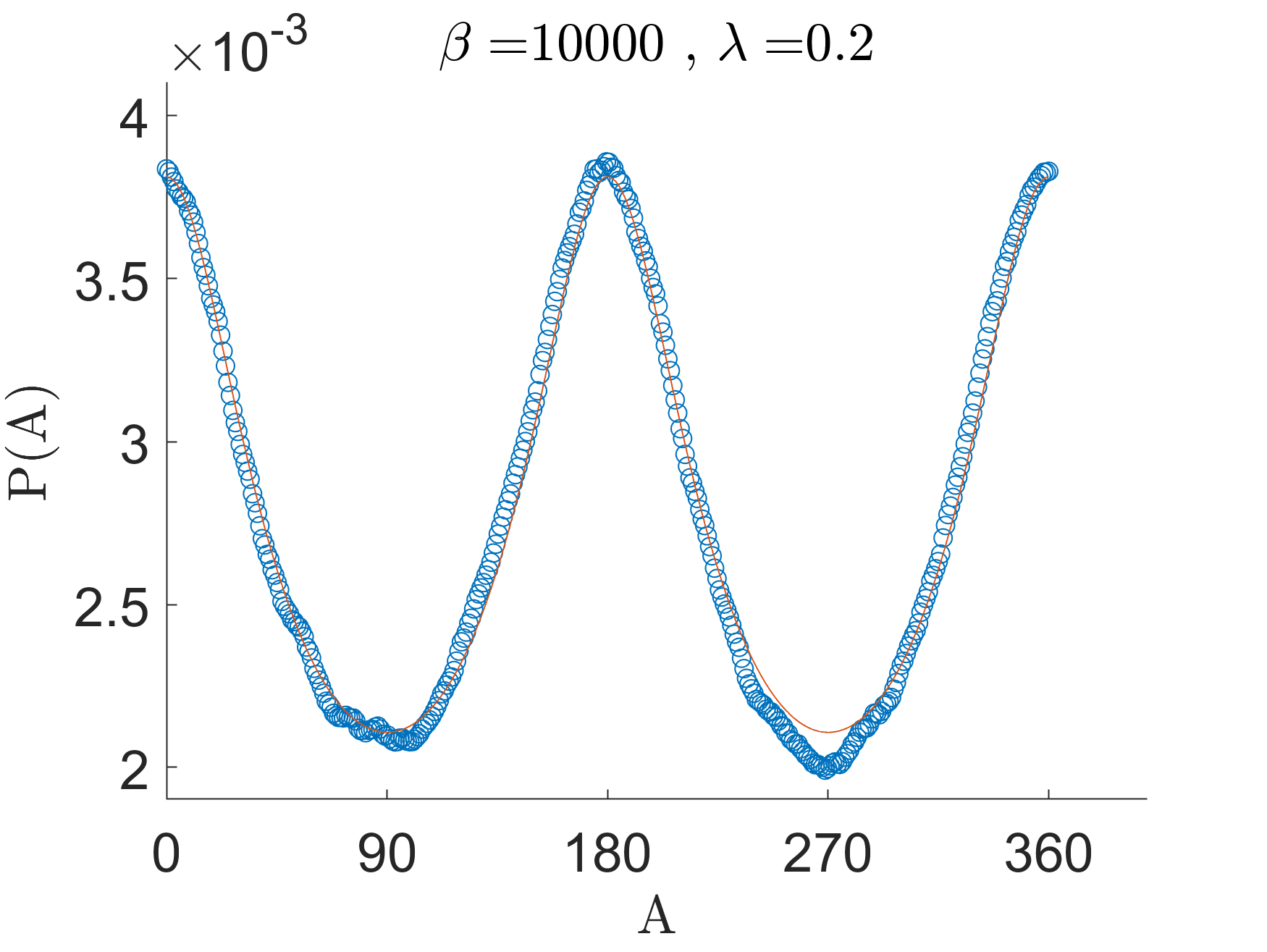}
    \includegraphics[width=2.3in]{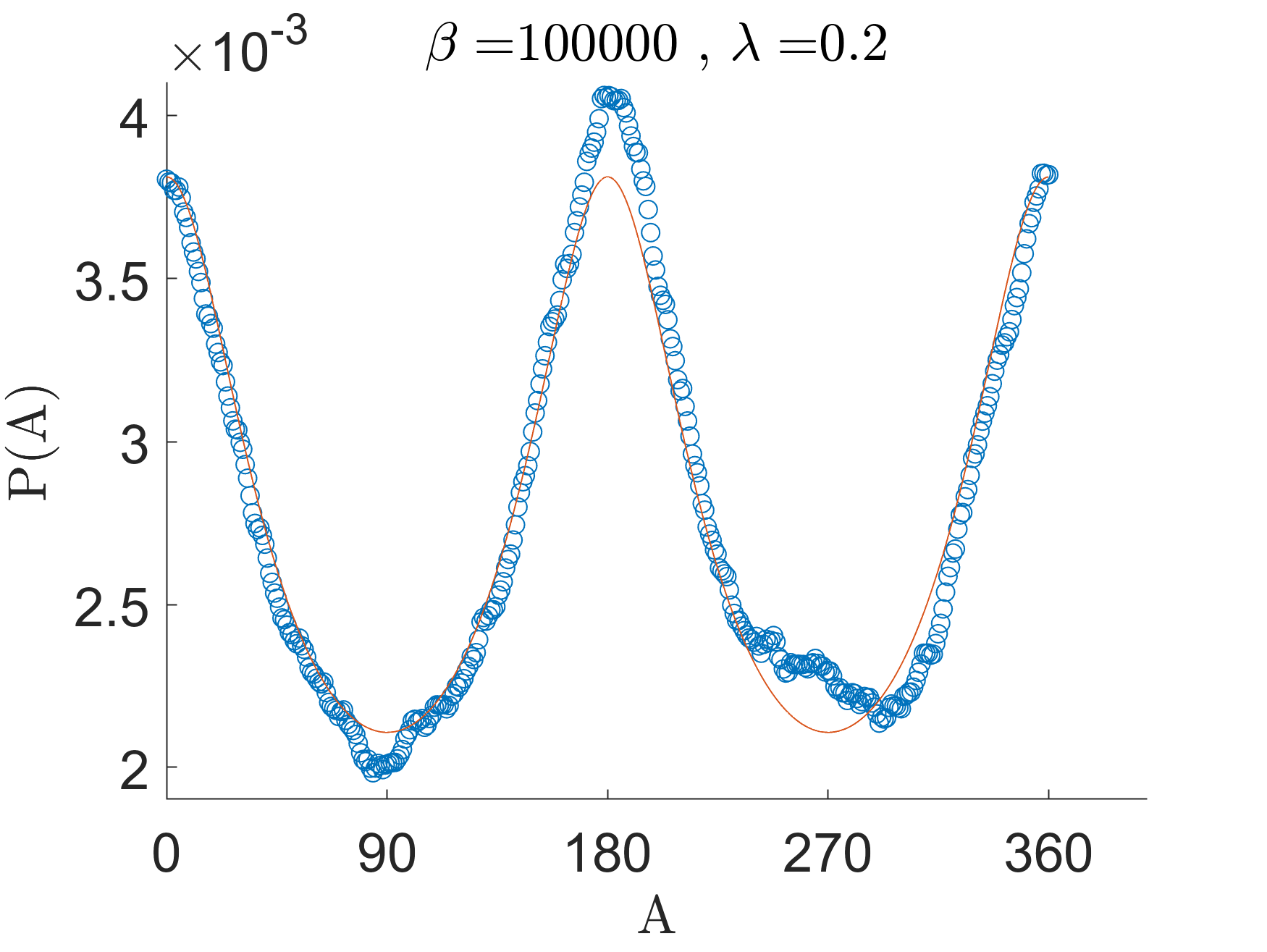}
\caption{Probability vs. $A$ (in degrees) obtained from Monte Carlo simulations of the asymmetric quadrumer. The plots are for the same anisotropy, $\lambda = 0.2$. They correspond to three values of inverse-temperature, $\beta = 5000$, $10000$ and $100000$ (left to right). In all three, data follow the same curve -- that of Eq.~\ref{eq.Pasym}. Deviations from the curve are comparable to error bars estimated from Monte Carlo runs.  } 
\label{fig.asymMC_fixlambda}
\end{figure*}

\begin{figure}
\includegraphics[width=\columnwidth]{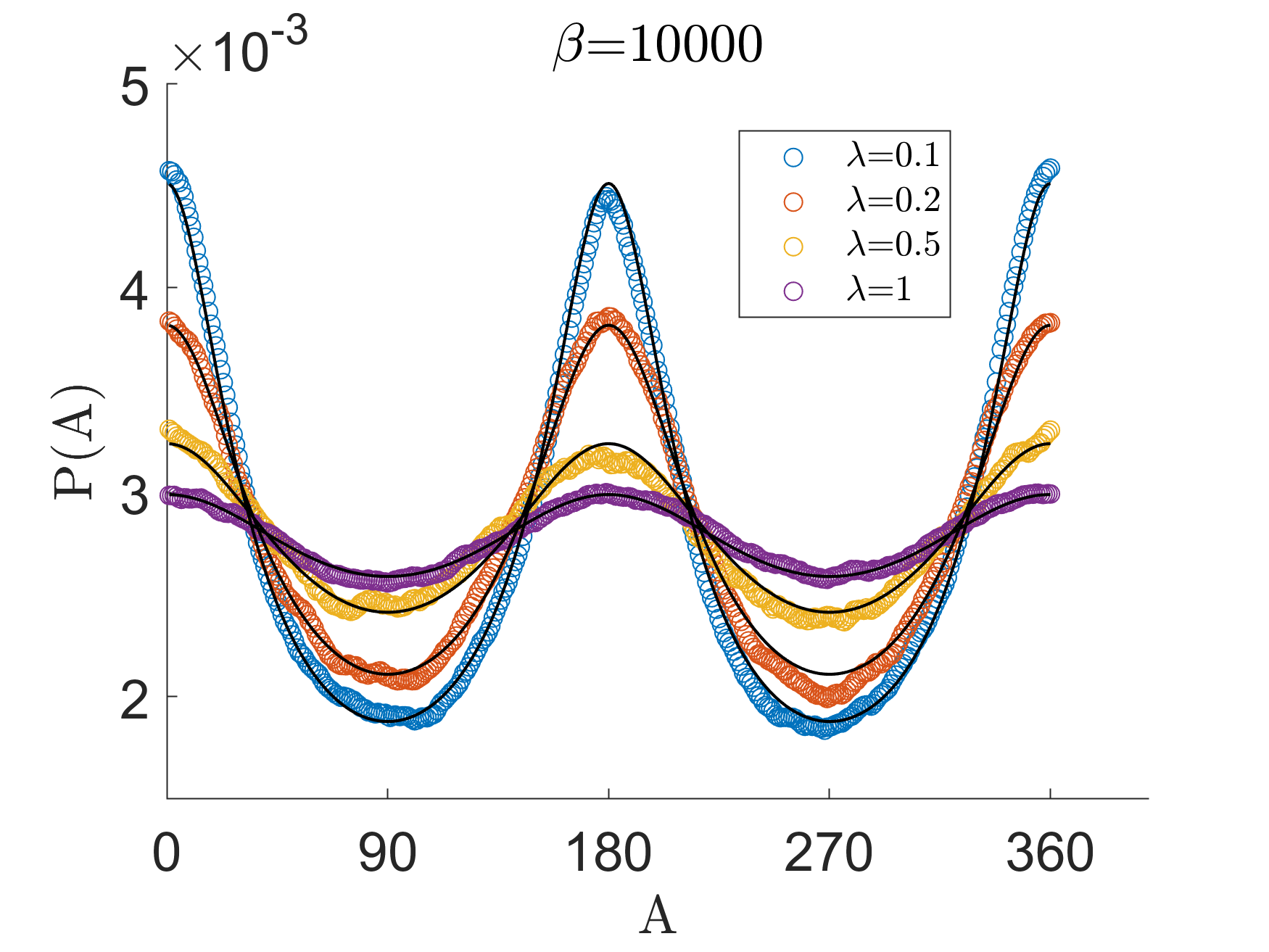}
\caption{Probability vs. $A$ (in degrees) for the asymmetric quadrumer obtained from Monte Carlo simulations. The plots are for the same temperature, $\beta = 10^4$. They correspond to four values of the anisotropy parameter: $\lambda = 0.1$, $0.2$, $0.5$ and $1$ (in order of decreasing spread in the vertical direction). For each $\lambda$, the data follow a curve obtained from Eq.~\ref{eq.Pasym} with the corresponding value of $\lambda$.  } 
\label{fig.asymMC_fixbeta}
\end{figure}

Fig.~\ref{fig.asymMC_fixlambda} shows the result for the asymmetric quadrumer with $\lambda = 0.2$. Probability is plotted against $A$ for three different values of $\beta=1/T$, the inverse temperature. Although $\beta$ varies over several orders of magnitude, the probabilities remain roughly the same. In all three plots, the data follow the same curve, given by Eq.~\ref{eq.Pasym}. Note that this curve depends on the value of the anisotropy parameter, $\lambda$, but not on $T$. Fig.~\ref{fig.asymMC_fixbeta} shows probabilities for four values of $\lambda$, but with $\beta$ held fixed. In all four, the data follow Eq.~\ref{eq.Pasym} with the corresponding value of $\lambda$.

These plots serve as numerical verification of the arguments in Sec.~\ref{sec.canonical} above. The asymmetric quadrumer was argued to show temperature-independent sampling. The degree of bias varies with $\lambda$, but not with temperature. These arguments were based on Eq.~\ref{eq.Hasymfluc}, with six quadratic terms in the energy. 
The equipartition theorem asserts that each quadratic term contributes a factor of $1/2$ to the specific heat. Indeed, Monte Carlo simulations yield specific heats close to 3 at low temperatures. 

\begin{figure*}
\includegraphics[width=2.3in]{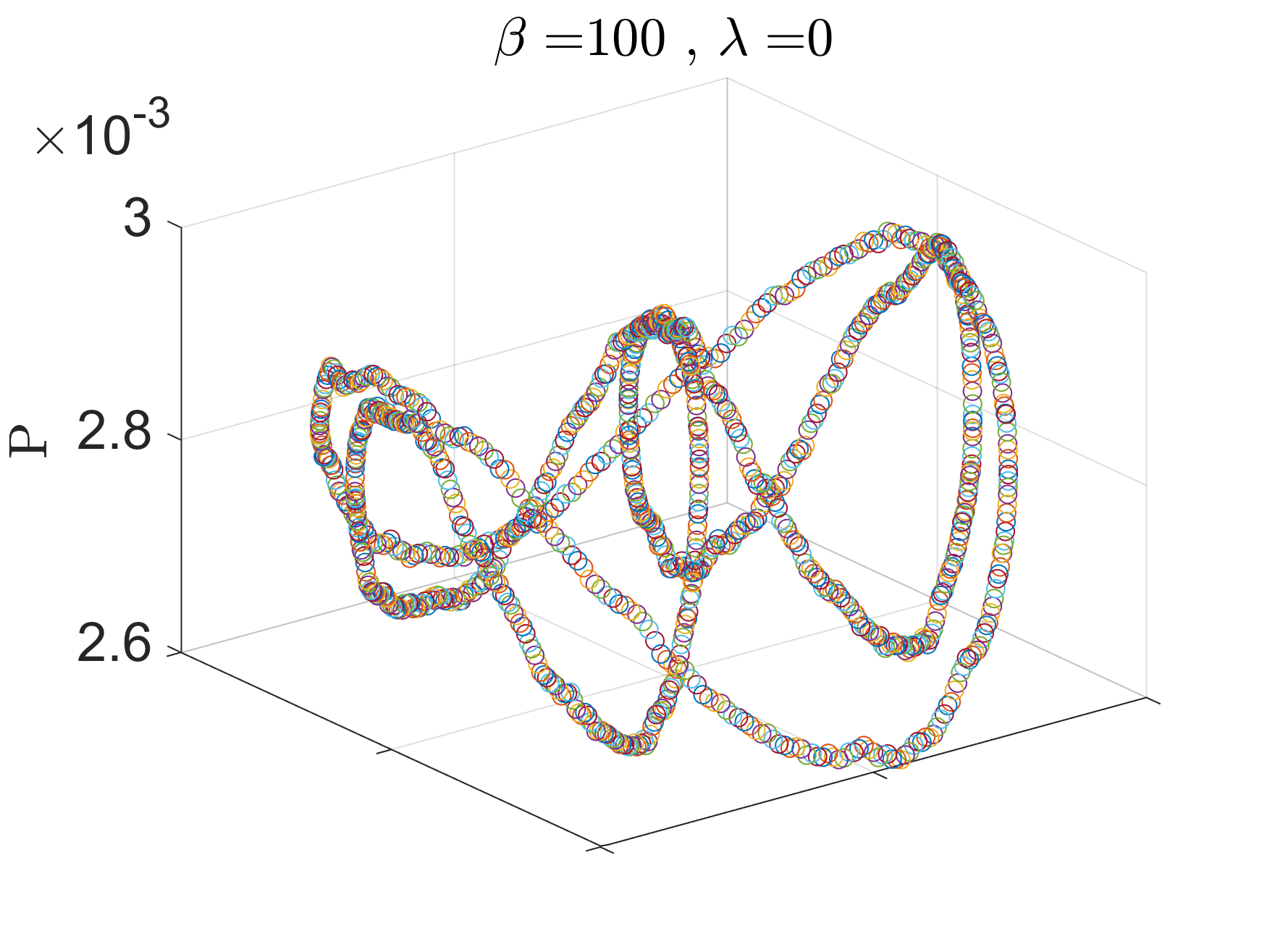}
\includegraphics[width=2.3in]{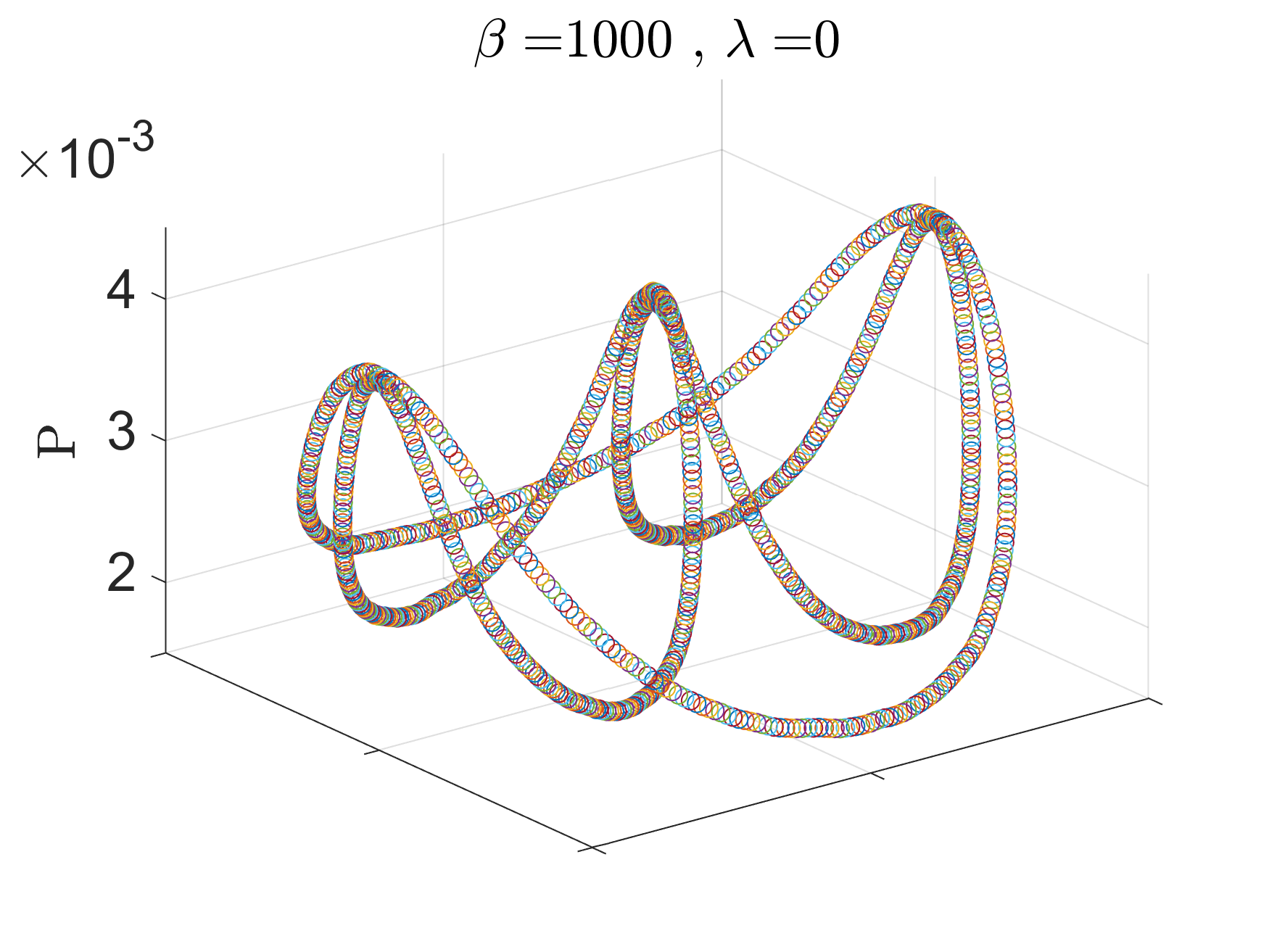}
\includegraphics[width=2.3in]{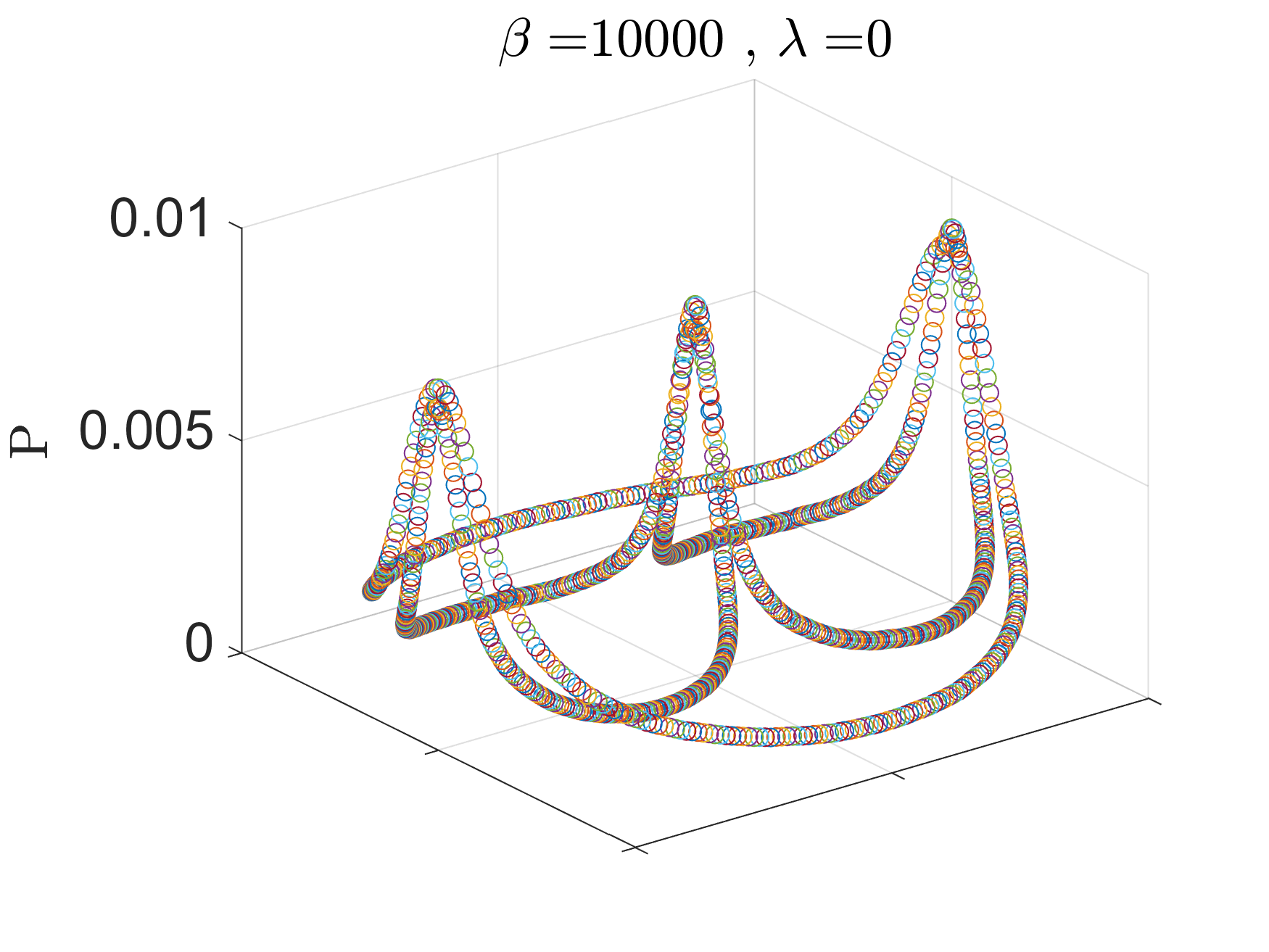}
\caption{Probability over the CGSS for the symmetric quadrumer, obtained from Monte Carlo simulations. The plots correspond to varying temperatures, $\beta=10^{2},10^{3},10^{4}$ (from left to right). Seen from the top, we have three circles as shown in Fig.~\ref{fig.cartoon}(b). The z-axis represents probability; note that the panels have different z-axis ranges. } 
\label{fig.symMC}
\end{figure*}

Monte Carlo results for the symmetric quadrumer are plotted in Fig.~\ref{fig.symMC}. Probability is plotted over the three circles of the CGSS, for three different values of $\beta$. For all three temperatures, we see that the collinear states (intersection points) have the highest probability. The lowest probability occurs for perpendicular states where the spin vectors point toward the corners of a square. As temperature is lowered, the probability profiles become sharper. The weight at intersection points (collinear states) grows dramatically. 

\begin{figure}
\includegraphics[width=\columnwidth]{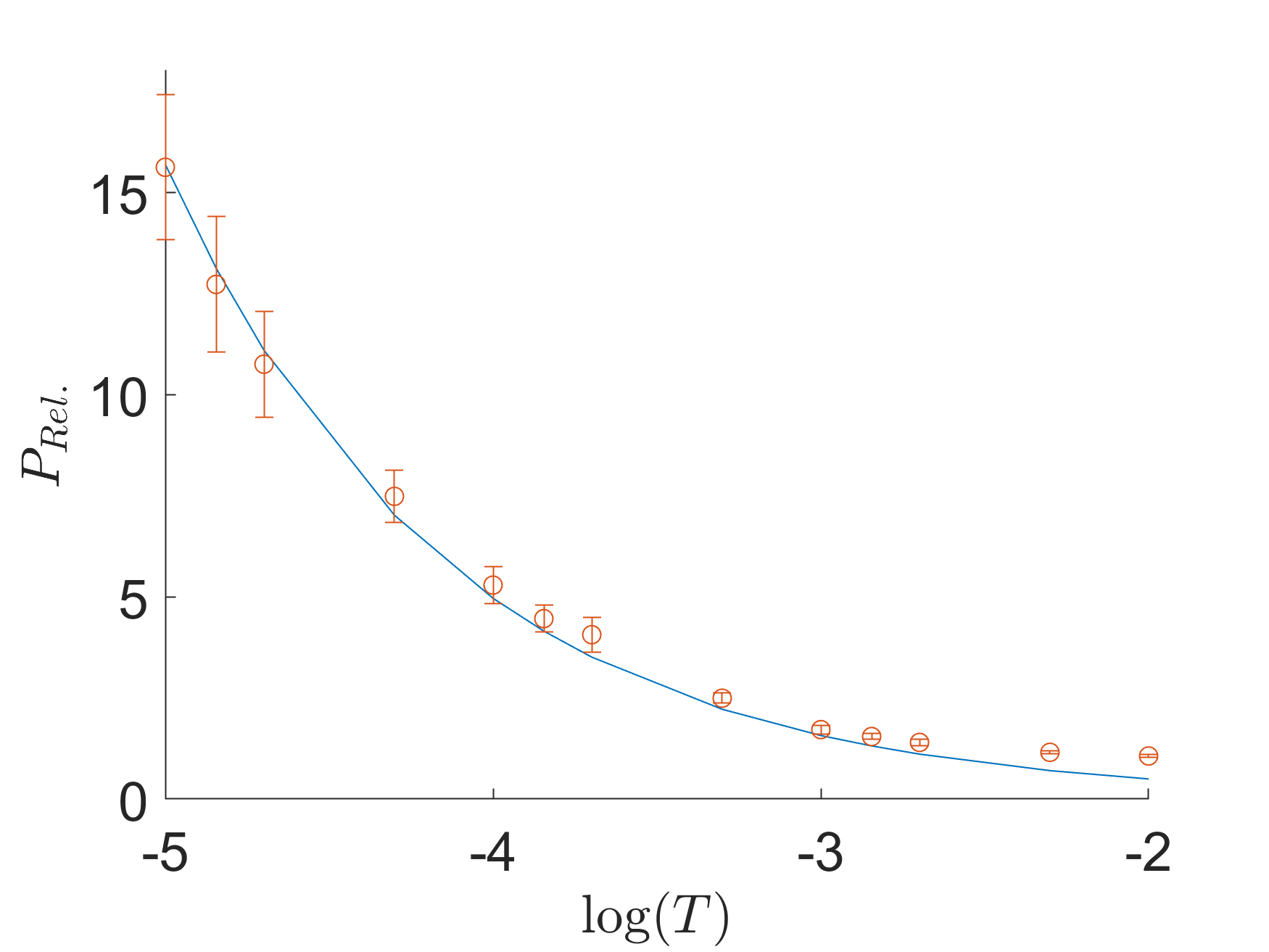}
\caption{Relative probability of collinear and perpendicular states in the symmetric quadrumer, obtained from Monte Carlo simulations. Error bars are estimated from the spread in values resulting from various choices of collinear and perpendicular states. The relative probability is plotted against $\log (T)$ (logarithm with base 10). The data is fit to $P_{rel.} \sim T^{-1/2}$. } 
\label{fig.relprobsym}
\end{figure}

This observation is quantified in Fig.~\ref{fig.relprobsym}. The x axis here represents temperature, while the y axis represents the relative probability of collinear and perpendicular states, $P(\mathrm{collinear})/P(\mathrm{perpendicular})$. Here, $P(\mathrm{collinear})$ is the sum of probabilities of all collinear states (representing three points of intersection on the CGSS space). As there are six perpendicular states in the CGSS, $P(\mathrm{perpendicular})$ is summed over them. This ratio grows dramatically as $T$ decreases. As shown in the figure, the temperature-dependence fits well to $\sim T^{-1/2}$. This verifies the arguments of Sec.~\ref{sec.canonical} regarding state-selection in the symmetric quadrumer. 

\section{Discussion} 

We have discussed sampling of classical ground states in two small clusters. Our conclusions may be extended to macroscopic magnets with frustration. The key requirements for our arguments are: (i) an extended space of classical ground states that does not result from symmetry, (ii) applicability of the ergodic and equiprobability hypotheses. 
These requirements can also be met in macroscopic systems. The honeycomb $J_1-J_2$ antiferromagnet provides an edifying example. Its CGSS is a contour in momentum space, with each point representing a spiral state\cite{Mulder_2010}. Order-by-disorder gives rise to a thermal phase transition where one spiral is `selected' by fluctuations\cite{Mulder_2010}. Above this critical temperature, a `ring liquid' phase appears\cite{Okumura2010} where all points on the CGSS are simultaneously sampled. Our results motivate further studies of the ring liquid phase. In particular, the sampling behaviour should be qualitatively different at $J_2/J_1=1/2$ -- a fine-tuned point where the CGSS contour self-intersects.

The honeycomb ring-liquid is one instance of a potentially large class of spin liquids, the `spiral liquids'\cite{Bergman_2007,Attig_2017,Niggemann_2020,Yao_2021,Yan_2022}.  
At least two material examples are known: MnSc$_2$S$_4$ and FeCl$_3$.  They show an extended peak in neutron scattering, which coincides with the CGSS contour expected on theoretical grounds\cite{Gao_2017,Gao2022}.
This can be interpreted as the system sampling the entire set of classical ground states. Studies so far have not carefully examined the sampling distribution or its evolution. There are limited results to show the existence of a phase that samples the entire CGSS, occurring above a critical ordering temperature, e.g., see Fig.~4 of Ref.~\onlinecite{Bergman_2007}. 
Our results suggest careful examination of sampling weights in neutron experiments as well as in simulations, particularly when the CGSS self-intersects.

We have demonstrated a qualitative difference between two classes of frustrated magnets. This difference originates from the topological character of the CGSS, distinguishing systems with a smooth manifold from those that self-intersect. 
Several materials and models are known in either class. Among materials, MnSc$_2$S$_4$\cite{Bergman_2007} is known to have a smooth CGSS while ErSn$_2$O$_7$\cite{Yan_2017} is proximate to a parameter regime with a self-intersecting CGSS. Among model systems, smooth CGSS' are found on the honeycomb \cite{Mulder_2010}, diamond\cite{Bergman_2007} and BCC\cite{Attig_2017} lattices as well as in the square Heisenberg-compass model\cite{Khatua_2023}. Self-intersections are found on the hyperhoneycomb\cite{Sungbin_2014} and HCP\cite{Niggemann_2020} lattices, as well as in the 1D Kitaev antiferromagnet\cite{Khatua_2021}.

In quantum spin clusters with self-intersecting CGSS', the geometry of intersections leads to bound-state-formation\cite{khatua_thesis,He_2023}. This constitutes a distinct mechanism for state selection, named order-by-singularity\cite{Khatua_2019}. Our results for the symmetric quadrumer can be viewed as a classical analogue of this phenomenon. At very low energies, the magnet is confined to the vicinity of an intersection point. In quantum magnets, this is due to bound-state-formation. In a classical setting, this is driven by enlarged phase space around singularities. The quantum bound-state-problem is strongly influenced by the co-dimension of intersections\cite{He_2023}. An interesting future direction is to explore the role of dimensionality in the classical problem.

\acknowledgments
We thank Han Yan, Johannes Reuther, Subhankar Khatua and Jeffrey Rau for insightful discussions. RG thanks the Natural Sciences and Engineering Research Council of Canada for support.

\appendix
\section{Low energy phase space}
\label{app.phasespace}
With four spins, phase space volumes are defined as
\begin{eqnarray}
\int  \Big\{\prod_{j=1}^4 d S_j^x d S_j^y d S_j^z~ \delta(\vec{S}_j \cdot \vec{S}_j - 1) \Big\}  g(\vec{S}_1,\ldots,\vec{S}_4),
\end{eqnarray} 
where $g(\vec{S}_1,\ldots,\vec{S}_4)$ is the sampling probability of a given neighbourhood. The $\delta$-functions enforce unit spin length. With twelve integration variables and four $\delta$-function constraints, the phase space volume is effectively eight dimensional. In the asymmetric quadrumer, low-energy configurations are described by Eq.~\ref{eq.forms} above. We have two degrees of freedom ($\phi_1$ and $\phi_2$) that select a particular ground state and six ($\ell_{1,2}$, $m_{1,2}$ and $\mu_{1,2}$) that encode fluctuations.
We change the integration variables to these new coordinates so that the phase space volume becomes 
\begin{eqnarray}
\nonumber \int \mathcal{J}(\phi_1,\phi_2,\ell_1,\ell_2,m_1,m_2,\mu_1,\mu_2) \times \\
\nonumber d\phi_1 ~d\phi_2 ~d\ell_1 ~d\ell_2 ~dm_1 ~dm_2 ~d\mu_1~ d\mu_2 \times \\
g(\phi_1,\phi_2,\ell_1,\ell_2,m_1,m_2,\mu_1,\mu_2),
\end{eqnarray}
where the Jacobian $ \mathcal{J}(\phi_1,\phi_2,\ell_1,\ell_2,m_1,m_2,\mu_1,\mu_2)$ evaluates to 32, up to corrections that are quadratic in fluctuation variables. 

We now consider the phase space of the microcanonical ensemble. The energy of the system is given by Eq.~\ref{eq.Hasymfluc}, with six quadratic terms. We rescale variables to set the coefficient of each of the six quadratic terms to unity. The accessible phase space then becomes a spherical shell in six dimensions with radius $\epsilon$ and thickness $d\epsilon$. This yields the phase-space-volume of Eq.~\ref{eq.vephi}.

\section{Finding the nearest point on the CGSS}
\label{app.nearest}
To find sampling probabilities from dynamics or Monte Carlo simulations, we assign a given configuration to a certain neighbourhood around the CGSS. Consider the asymmetric quadrumer where the CGSS is a circle. Given a low-energy configuration, we express it in the form of Eq.~\ref{eq.forms}. For example, we identify the in-plane component of $(\vec{S}_1 - \vec{S}_2)/2$ as $\hat{n}(\phi_1)$ and that of $(\vec{S}_3 - \vec{S}_4)/2$ as $\hat{n}(\phi_2)$. Having thus extracted $\phi_1$ and $\phi_2$, we identify the CGSS coordinate as $A = \phi_2 - \phi_1$ in accordance with Fig.~\ref{fig.cartoon}.

In the symmetric quadrumer, we have an additional layer of complexity. As the CGSS has three circles, we must first assign a given configuration to one of the circles. To do so, we consider three vector quantities: $\vec{S}_1 - \vec{S}_2$, $\vec{S}_1 - \vec{S}_3$ and $\vec{S}_1 - \vec{S}_4$. Based on which of these three has the largest magnitude, we identify the nearest circle. We then proceed in the same way as with the asymmetric quadrumer to find the angle coordinate ($A$, $B$ or $C$). If two of these vectors are comparably large (with magnitude close to 2), the configuration is close to an intersection point.

\bibliographystyle{apsrev4-1} 
\bibliography{ObS_classical}
\end{document}